\documentclass[aps,reprint,twocolumn,showpacs,preprintnumbers,amsmath,amssymb,nofootinbib,superscriptaddress,showkeys]{revtex4-1}

\usepackage{epsfig}
\usepackage{slashed}

\usepackage{color}

\begin{document}

\title{Power Counting of Contact-Range Currents in Effective Field Theory}
	\author{M. Pav\'on Valderrama}\email{pavonvalderrama@ipno.in2p3.fr} 
\affiliation{Institut de Physique Nucl\'eaire, Universit\'e Paris-Sud, IN2P3/CNRS, F-91406 Orsay Cedex, France}
        \author{D.R. Phillips}\email{phillips@phy.ohiou.edu}
\affiliation{Institute of Nuclear and Particle Physics and 
Department of Physics and Astronomy, Ohio University, Athens, Ohio 45701, USA}
\date{\today}

\begin{abstract} 
\rule{0ex}{3ex} 
We analyze the power counting of two-body currents in nuclear effective
field theories (EFTs). 
We find that the existence of non-perturbative physics at low energies, which
is manifest in the existence of the deuteron and the ${}^1$S$_0$
NN virtual bound state,
combined with the appearance of singular potentials in versions of nuclear EFT that incorporate chiral symmetry,
modifies the renormalization-group flow of the couplings associated with
contact operators that involve nucleon-nucleon pairs and external fields.
The order of these couplings is thereby enhanced with respect to the naive-dimensional-analysis
estimate. Consequently,  short-range currents enter at a lower order in the 
chiral EFT than has been appreciated up until now, and their
impact on low-energy observables is concomitantly larger. 
We illustrate the changes in the power counting with a few low-energy processes
involving external probes and the few-nucleon systems, including
electron-deuteron elastic scattering and
radiative neutron capture by protons.
\end{abstract}


\maketitle

Effective field theories (EFTs) describe physics at low momenta ---
specified by the soft scale $Q$ --- where the fields and symmetries out of
which they are constructed are well-defined and make sense. $Q$ is
small in comparison to the natural ultraviolet (UV) cutoff of the EFT,
the hard scale $M$, which corresponds to the energy region where the
EFT's degrees of freedom no longer describe the physics. 
As long as $Q < M$, the EFT provides an expansion of
observables in powers of $Q/M$.
To avoid explicit sensitivity to the physics at the hard scale $M$,
EFTs are regularized and renormalized.
In the Wilsonian formulation~\cite{Wilson:1973jj} we regularize by means
of a UV cutoff $\Lambda$ that serves as an explicit separation
between low- and high-energy physics.
The UV cutoff is not the natural cutoff $M$ but rather a theoretical device
for analyzing the EFT and we must make sure that calculations
do not depend on $\Lambda$, i.e. we renormalize the theory.
By reducing the cutoff from $M$ to $Q$ we can analyze the evolution
of the EFT couplings and determine their relative importance
at low energies.

The Wilsonian renormalization group (RG) is a well-established tool
in the context of EFTs where the expansion is strictly
perturbative~\cite{Polchinski:1983gv} ---the Standard Model,
Chiral Perturbation Theory ($\chi$PT), QED below the weak scale---
and helps to explain why these theories work~\cite{Lepage:1989hf}.
However, over the last twenty years an EFT has been developed
for nuclear physics,
in which already at leading-order (LO) the EFT two-nucleon
amplitude contains poles (bound states) and
the potentials are singular,
behaving as $1/r^3$ as $r \rightarrow 0$.
This  provokes novel questions about the meaning of renormalization
in this context, questions that have led to much
controversy~\cite{Nogga:2005hy,PavonValderrama:2005wv,Birse:2005um,Epelbaum:2006pt,Epelbaum:2009sd}
and new approaches~\cite{Long:2007vp,Valderrama:2009ei,Valderrama:2011mv,Long:2011qx,Long:2011xw,Long:2012ve}.
In this paper we examine the matrix elements of electroweak current
operators in nuclear EFTs. We show how RG invariance can be used to
determine the order at which these operators enter the EFT expansion
for the electromagnetic or weak-nuclear currents by which nuclei
couple to electrons, photons, and neutrinos.
We demonstrate that
naive dimensional analysis (NDA) underestimates the role
of these current operators; in essence because it neglects their
anomalous dimension.

This has significant implications for the theory of processes including
elastic and inelastic electron-deuteron and electron-trinucleon
scattering~\cite{Phillips:2006im,Piarulli:2012bn}, the proton fusion reaction $pp \rightarrow d e^+ \nu_e$~\cite{Marcucci:2013tda}
and muon capture on deuterium and ${}^3$He~\cite{Marcucci:2011jm,Gazit:2008vm}.  Until now the most 
sophisticated nuclear EFT calculations of these processes have invoked NDA, since all have employed
$\chi$PT power counting to organize nuclear operators. (See Ref.~\cite{Epelbaum:2008ga} for a recent review.) 
 The results of Refs.~\cite{Phillips:2006im,Piarulli:2012bn,Marcucci:2013tda,Marcucci:2011jm,Gazit:2008vm} have implications for the structure
of light nuclei, solar models, and precision tests of the Standard
Model.  Here we argue that short-distance contributions to the
current operators used therein are actually significantly more
important than was appreciated in these works.  In particular, we show
that RG invariance requires that in most Gamow-Teller or M1
transitions in few-nucleon systems the short-distance contribution
enters at least one order earlier than is predicted in the $\chi$PT
power counting originally suggested
by Weinberg~\cite{Weinberg:1990rz,Weinberg:1991um}.

In EFT observable quantities do not depend on the choice of the
cutoff. This can be realized by imposing the cutoff independence of
(here, on-shell) matrix elements:
\begin{eqnarray}
\frac{d}{d \Lambda}\,\langle \Psi' | \mathcal{O}_{\rm EFT} | \Psi \rangle
= 0 \, , \label{eq:cutoff-full}
\end{eqnarray}
with $\mathcal{O}_{\rm EFT}$ an EFT operator, and $\Psi$ ($\Psi'$)
the initial- (final-) state EFT wave functions.
For concreteness we assume that $\mathcal{O}_{\rm EFT}$ is a component
of a nuclear (four-) current, which depends on the momentum of the probe
$\mathcal{O}_{\rm EFT} = \mathcal{O}_{\rm EFT}({\bf q})$.
Note that Refs.~\cite{Nakamura:2006hc,Kvinikhidze:2007eu} already formulated 
RGs akin to Eq.~(\ref{eq:cutoff-full}) for NN matrix elements of $\mathcal{O}_{\rm EFT}$, 
although no conclusions 
regarding the chiral EFT ($\chi$EFT) power counting were drawn.
See Refs.~\cite{Birse:1998dk,Barford:2002je,Birse:2005um} for applications
 to this RG of the two-nucleon potential.

$\chi$EFT operators contain one-body, two-body, three-body,
etc. contributions---separated according to how many nucleons
participate directly in the interaction with the external
probe. Two-body operators can be subdivided into pion-range and contact parts 
(but see also below):
\begin{eqnarray}
\mathcal{O}_{\rm EFT} =
\mathcal{O}_{\rm 1B} +  \mathcal{O}_{{\rm 2B}, \pi} +
\mathcal{O}_{{\rm 2B}, C} + \ldots\, .
\end{eqnarray}
In what follows we focus on the two-body operators, which in general
are the dominant correction to the one-body piece. The results can be
generalized to higher-body current operators.
For $\mathcal{O}_{{\rm 2B}, \pi}$ the interaction among nucleons is
mediated by pions.
In general the power counting of this piece is straightforward: we
simply count the powers of $Q \equiv {\bf q}, m_\pi, {\bf p}$ (with
$m_\pi$ the pion mass and ${\bf p}$ any nucleon momenta on which
${\cal O}$ depends) in each piece of the operator, and assume they are
made up by powers of the breakdown scale $M$ in the coefficient of
that part of ${\cal O}_{\rm EFT}$.
In other words we assume naive dimensional analysis (NDA).
This leads straightforwardly to the conclusion that
${\cal O}_{{\rm 2B},\pi}$ is typically suppressed relative to the one-body
contribution, as first articulated in Refs.~\cite{Weinberg:1992yk,Friar:1996zw}.
It also produces a ${\cal O}_{{\rm 2B},\pi}$ which---up to 
contact-term pieces---operates at a range $r \sim 1/m_\pi$.

If ${\cal O}_{{\rm 2B},\pi}$ has divergent parts those will appear
in the final answer as contributions of contact range. They will then
depend on a regularization scale. But that regularization
scale can be kept distinct from the scale $\Lambda$ used
to regularize the Schr\"odinger equation. Furthermore, such 
contact pieces of ${\cal O}_{{\rm 2B},\pi}$ will have at least the NDA
order of $\mathcal{O}_{{\rm 2B}, C}$. They cannot produce
the enhanced-over-NDA contact-range currents that are our 
concern here. 

Power counting for $\mathcal{O}_{{\rm 2B}, C}$ is more subtle,
but  a few simplifications help us determine it.
First, we only have to consider the leading-order piece of
this part of the operator: subleading contact-range currents
are trivially suppressed by the extra powers of $Q$
contained in the operators.
Second, ${\cal O}_{{\rm 2B},C}$ cancels the cut-off dependence of
${\cal O}_{{\rm 1B}}$ and ${\cal O}_{{\rm 2B},\pi}$, yet the two-body
piece is suppressed with respect to the one-body.
Thus we can simply ignore ${\cal O}_{{\rm 2B},\pi}$ at lowest order.
This is still true even if we promote the pion-exchange currents
by one order (as would happen if we adopted the power counting
that justifies the iteration of one-pion exchange in the leading-order
NN potential in Ref.~\cite{Birse:2005um}).
This yields the following RG equation for the leading piece
of $\mathcal{O}_{{\rm 2B}, C}$:
\begin{eqnarray}
\frac{d}{d \Lambda}\,\langle \Psi' | \mathcal{O}_{{\rm 2B}, C}^{(0)} | \Psi \rangle
&=&  - \frac{d}{d \Lambda}\,\langle \Psi' | \mathcal{O}_{\rm 1B}^{(0)}| \Psi \rangle +  \dots \, , 
\label{eq:cutoff-C}
\end{eqnarray}
where the dots indicate the higher-order terms.
The leading $\mathcal{O}_{{\rm 2B}, C}$ most often contains no powers
of the small scale (but see the example of the charge operator, below)
so here we write it schematically as $ \mathcal{O}_{{\rm 2B},
C}^{(0)}=C_0(\Lambda) \delta^{(6)}_\Lambda(r', r)$, where the
subscript indicates that the $\delta$ function is also regulated at
scale $\Lambda$. The quantum numbers of the current will be carried by
an operator which we have not written here.

The (leading) renormalization-group invariance of the current matrix
element is now encoded in a differential equation for $C_0(\Lambda)$,
which is:
\begin{eqnarray}
 \frac{d}{d \Lambda} \left[ C_0(\Lambda)\,
\langle \Psi'|\delta_\Lambda^{(6)}| \Psi\rangle \right] = 
 - \frac{d}{d \Lambda}\,
\langle \Psi' | \mathcal{O}_{\rm 1B}^{(0)}| \Psi \rangle.
 \label{eq:RGE}
\end{eqnarray}
This is an inhomogeneous first-order differential equation, although in practice we can ignore the right-hand side.
Given a boundary condition it has a unique solution, which determines
the power counting of $\mathcal{O}_{{\rm 2B}, C}$ and, concomitantly,
its matrix elements.

The boundary condition results from the observation that
if we set the cutoff to be the size of
the natural cutoff of the theory, i.e. $\Lambda = M$,
then  $C_0$ can only scale with $M$.
We take $C_0$ to be a coupling of inverse mass dimension $d$, where that
dimension is determined by the particular operator it multiplies,
and so we have $C_0(M) \sim M^{-d}$. This is NDA applied at the scale $M$.
In many EFT applications one is interested in estimating the
size of $C_0(\Lambda)$ prior to any examination of data, and this
naturalness assumption provides a way forward without which the EFT
power counting cannot be determined.

Once this boundary condition is chosen Eq.~(\ref{eq:RGE}) determines
$C_0(\Lambda)$ for any $\Lambda$. Although the equation can be integrated in two
directions, the standard practice is to evolve the couplings from
$\Lambda \sim M$ to $\Lambda \sim Q$ to find out how integrating
high-momentum modes out of the theory affects the size of the EFT
operators that must compensate for their removal.  In this view the
EFT results from infrared RG evolution of a more fundamental
theory. In practice $\chi$EFT calculations are carried out with a
cutoff $\Lambda$ that lies in between the high-energy scale of
$\chi$PT, $M \approx 1$ GeV, and the soft scales $Q \approx 150$
MeV. Since the operators in question often have large inverse mass
dimension understanding their running from $M$ to these lower scales
has significant practical importance.

Now, if the wave functions $|\Psi \rangle$ and $|\Psi' \rangle$ are
plane waves, then the matrix elements appearing in Eq.~(\ref{eq:RGE})
have no dependence on $\Lambda$ and it reduces to
$\frac{dC_0(\Lambda)}{d\Lambda}=0$.  This, together with the
assumption that $C_0(M)$ is natural with respect to the scale $M$,
means that $C_0(\Lambda) \sim M^{-d}$ for all $\Lambda$. This is the
power counting on which $\chi$EFT for few-nucleon systems has been
based. Such a power counting is valid if the solutions for nuclear
wave functions are plane waves, or reduce to plane waves in the
UV. (For example, the sub-leading corrections to naive dimensional analysis
that are present in the UV in the case of Coulombic wave functions do not
alter the LO results that we derive here.)

However, if $| \Psi \rangle$ and $| \Psi' \rangle$
do not behave like plane waves at momenta $\gg Q$, then in general we will have
$\langle \Psi'|\delta_\Lambda^{(6)}| \Psi\rangle  \sim \Lambda^{a}$,
with the dimensions of the matrix element made up by soft scales $Q$. 
Using Eq.~(\ref{eq:RGE}) to evolve the value of $C_0(\Lambda)$ to the
soft scale $Q$ we find the infrared enhancement (p.v. $a>0$):
\begin{eqnarray}
C_0(\Lambda \sim Q) \sim \frac{1}{M^{d-a} Q^a} \, .
\end{eqnarray}
Consequently the contact-range operator receives a promotion of $a$
orders in the EFT expansion of ${\cal O}$ in powers of
$Q/M$. $-a$ is the anomalous dimension of $C_0$ in this EFT and
is non-zero because of the strong interactions in the nuclear wave
functions. The value of $a$ is independent of how many powers of the
soft scales ${\bf q}$ and $m_\pi$ the operator carries, and so
is the same for higher-order coefficients in the expansion of ${\cal O}_{{\rm 2B}, C}$
in powers of $Q$.
For instance, nothing in the above analysis changes if the operator is
$\langle {\bf p}'|\vec{\mathcal{O}}_C ({\bf q}; \Lambda) | {\bf p}
\rangle = M(\Lambda)\, {\bf \beta} \times {\bf q}$, with $\vec{\beta}$
an arbitrary vector.
The coefficient $M(\Lambda)$ will then also be
enhanced by the factor $1/Q^a$ with respect to the NDA result.

The power counting enhancement can be determined by examining the
behavior of the distorted wave functions $| \Psi \rangle$, $| \Psi'
\rangle$ at short distances. In order to demonstrate this we choose a
specific regularization of $\delta^{(6)}$:
\begin{eqnarray}
\langle \vec{r}' | \mathcal{O}_{{\rm 2B},C}^{(0)} | \vec{r} \rangle=
C_0(R)\,\frac{\delta(r - R)}{R^2}\,\frac{\delta(r' - R)}{R^2}\,
X(\hat{r}, \hat{r}') \, ,
\end{eqnarray}
with $X$ referring to the non-radial piece of the operator.  Here we
have introduced a co-ordinate space cutoff $R$ that is related to the
$\Lambda$ of the previous paragraphs via $R \propto 1/\Lambda$. The
proportionality constant does not affect the value of $a$. 

Similarly we are not interested here in the numerical factor (reduced
matrix element) generated by the matrix element of
$X(\hat{r},\hat{r}')$ between the angular pieces of the nuclear wave
functions. If we write them as:
\begin{eqnarray}
\Psi'(\vec{r}\,') = \frac{u'(r)}{r}\,Y'(\hat{r}\,') \, , \quad
\Psi(\vec{r}\,) = \frac{u(r)}{r}\,Y(\hat{r}) \, , 
\end{eqnarray}
with $Y'$ and $Y$ non-radial pieces containing the dependence on
angular momentum and other unspecified quantum numbers, then the
matrix element yields:
\begin{eqnarray}
\langle \Psi' | \delta^{(6)}_{\Lambda} | \Psi \rangle \propto
\frac{u(R)}{R}\,\frac{u'(R)}{R} \, .
\end{eqnarray}
Thus all we need to know is the behavior of the wave at $R \ll 1/Q$ to
get the anomalous dimension of $C_0(R)$. In particular, if we
assume that 
\begin{equation}
u(R)/R \sim R^{b}; \quad u'(R)/R \sim R^{c},
\end{equation}
the anomalous dimension is $-a = b + c$.
The problem of determining the power counting for the leading
contact-operator contribution to nuclear currents is thus reduced to
the simple matter of computing the UV spectral indices $b$ and $c$ of
the EFT wave functions.  Here we analyze only the leading contact
operator, so wave functions computed at LO in $\chi$EFT are adequate
for this purpose.

%------------------------------------------------------------------------
\begin{table*}[tb]
\caption{\label{tab:counting}
Power counting of contact-range currents for some observables of
interest involving the deuteron and/or ${}^1$S$_0$ NN state.
(Note that the order given for the dominant one-body effect
pertains to $\chi$EFT, and not to the pionless EFT.)
The reactions we are considering are electron-deuteron scattering,
radiative neutron capture by protons and proton-proton fusion.
The observables we list are: the squared deuteron electromagnetic
radius ($r_{\rm em}^2$), the deuteron magnetic dipole ($\mu_d$) and
electric quadrupole moments ($Q_d$) and the $M1$ matrix element
for neutron capture and proton-proton fusion
We also list the schematic form of the lowest order two-body contact-range
current operator that contributes to each one of these observables, expressed
in the plane wave basis, where ${\bf q}$ is the momentum
of the external probe.
}
\begin{ruledtabular}
\begin{tabular}{c | c | c | c | c | c | c }
      Process & Matrix Element & $1B$ & $2B$ ($\mathcal{O}_C$)
      & $2B$(NDA) & $2B$($\slashed{\pi}$) & $2B$($\pi$)\\
      \hline
      $d e \to d e$ & $r_{\rm em}^2$ & $\rm LO$ ($Q^0$) &
      $D(\Lambda)\,{{\bf q}\,}^2$ & $\rm N^5LO$ ($Q^5$) & $\rm N^3LO$ ($Q^3$) &
      $\rm N^{9/2}LO$ ($Q^{9/2}$) \\
      & $\mu_d$ & $\rm LO$ ($Q^{1}$) &
      $M(\Lambda)\,\vec{\beta} \times {\bf q}$ & $\rm N^3LO$ ($Q^4$)
      & $\rm NLO$ ($Q^2$) &
      $\rm N^{5/2} LO$ ($Q^{7/2}$) \\
      & $Q_d$ & $\rm LO$ ($Q^0$) & $Q(\Lambda)\,T_2({\bf q})$ &
      $\rm N^3LO$ ($Q^5$) & $\rm NLO$ ($Q^3$) &
      $\rm N^{5/2}LO$ ($Q^{4.5}$)\\ 
      \hline
      $np \to d\gamma$ & $M1$ & $\rm LO$ ($Q^1$) &
      $M(\Lambda)\,\vec{\beta} \times {\bf q}$ & $\rm N^3LO$ ($Q^4$)
      & $\rm NLO$ ($Q^2$) & $\rm N^{7/4}LO$ ($Q^{11/4}$) \\
      \hline
      $pp \to d e^{-} \bar{\nu}_e$ & $M1$ & $\rm LO$ ($Q^{0}$) &
      $A(\Lambda)\,\vec{\beta}$ & $\rm N^3LO$ ($Q^3$)
      & $\rm NLO$ ($Q^1$) & $\rm N^{7/4}LO$ ($Q^{7/4}$) \\
   \end{tabular}
\end{ruledtabular}
\end{table*}
%------------------------------------------------------------------------

We now illustrate these ideas by examining a few processes
involving the nucleon-nucleon (NN) system. 
For concreteness we will begin by considering electromagnetic reactions
where the NN system interacts with a (real or virtual) photon. 
Consequently, the current operator has a Lorentz index
and must fulfil the Ward identity (i.e. the continuity equation): 
\begin{eqnarray}
\langle J^{\mu}({\bf q}) \rangle =
\langle \Psi' | \mathcal{O}^{\mu}({\bf q}) | \Psi \rangle \, , \quad
q_{\mu}\,\langle J^{\mu}({\bf q}) \rangle = 0 \, ,
\label{eq:WI}
\end{eqnarray}
a constraint that has consequences for the contact-range
currents in the charge (i.e. $\mu = 0$) form factor.
That quantity is defined as
\begin{eqnarray}
|e| G_C({\bf q}) = \overline{\langle \Psi_d | J^{0}({\bf q}) | \Psi_d \rangle}
\end{eqnarray}
where $| \Psi_d \rangle$ is the deuteron wave function and the bar
indicates averaging over spins.
The LO operator that contributes to the charge form factor
is the one-body charge, which in the plane-wave basis reads
$\langle {\bf p}'| J^0_{1B, LO}({\bf q})| {\bf p} \rangle =
|e|\,\delta^3(p'-p-q)$. It yields a $G_C$ of order $e Q^0$. 
Now, Eq.~(\ref{eq:WI}) implies that $G_C({\bf 0}) = 1$, but, as long as an
energy-independent potential generates $|\Psi_d \rangle$, and 
 $\langle \Psi_d|\Psi_d \rangle=1$, then $\overline{\langle
  \Psi_d | J^0_{1B, LO}({\bf 0}) | \Psi_d \rangle}=1$. Consequently
all higher-order contributions to the charge operator must vanish at
${\bf q}=0$.  The pion-range current operator will satisfy this
requirement if constructed using dimensional regularization and a
mass-independent renormalization scheme (as in
Refs.~\cite{Kolling:2009iq,Kolling:2011mt,Pastore:2008ui,Pastore:2011ip}), and so the lowest-order, non-trivial,
contact operator that contributes to $G_C$ is $D(\Lambda)\,{{\bf
    q}\,}^2$. This can be thought of as a short-distance contribution
to the deuteron's charge radius. According to NDA it affects $G_C$ at
$O(Q^5)$.
However, this does not take into account the anomalous dimensions stemming
from the wave functions.

In $\chi$EFT the LO nucleon-nucleon (NN) potential behaves like
$1/r^3$ plus a delta-function at short distances. The singular
potential is renormalized by the delta function, but results in wave
functions $u$ and $w$ $\sim r^{3/4}$
as $r \rightarrow 0$~\cite{PavonValderrama:2005uj,PavonValderrama:2005gu}.
The power-law exponents in the ${}^3$S$_1$-${}^3$D$_1$ NN partial wave are
thus $b = c = -1/4$.  This makes the contribution of
$D(\Lambda)\,{{\bf q}\,}^2$ slightly bigger: it enters $J^0$ at $O(e
Q^{4.5})$. 

If we consider the deuteron magnetic and quadrupole form factors, the
lowest order contact operators take the schematic form
$M(\Lambda)\,\vec{\beta} \times {\bf q}$ and $Q(\Lambda)\,T_2({\bf
  q})$ (with $T_2$ a tensor involving two powers of ${\bf
  q}$). According to NDA they appear at $O(Q^4)$ and $O(Q^5)$
respectively.  As discussed above, their tensor structure does not
affect our RG argument, so they too receive a slight enhancement, to
$O(Q^{7/2})$, and $O(Q^{9/2})$, respectively.  We note that
the magnetic form factor starts only at $O(Q)$, and the relative
importance of the chiral EFT short-distance contribution there may
explain the difficulties of some models to reproduce the deuteron
magnetic moment (see, e.g., the discussion in Ref.~\cite{Arenhovel:1999nq}). The
enhancement of the short-distance part of the quadrupole operator
strengthens the argument of Ref.~\cite{Phillips:2006im} that this operator is key
to accurate description of $G_Q({\bf q})$ in the low-${\bf q}^2$
regime.

For comparison, in the ``pionless EFT", the NN
potential that generates $|\Psi_d \rangle$ operates strictly at $r=0$, and so
we have $b=c=-1$. Our analysis then  reproduces  the well-developed power counting of
electromagnetic operators in pionless EFT~\cite{Kaplan:1998sz,Chen:1999tn,Beane:2000fx,Bedaque:2002mn}. We include these results in
Table~\ref{tab:counting} but do not discuss them further. (The pionless EFT power counting in the strong sector is derived from the Wilsonian RG 
in Ref.~\cite{Harada:2006cw}.) 

For radiative capture of neutrons by protons (or, equivalently,
photodisintegration of the deuteron) at threshold the M1 transition
dominates, and the momentum structure of the operator is as for the
deuteron magnetic moment.
The only difference is that the incoming NN partial wave is
${}^1$S$_0$ while the outgoing state is still a deuteron. This is,
however, important, since the LO $\chi$EFT wave function in the
${}^1$S$_0$ behaves as $1/r$ at short distances, i.e. we have $b=-1$.
Physically this occurs because one-pion exchange, which is not a
singular potential for spin-0 partial waves, is too weak to generate
the unnaturally large scattering lengths in this channel ($a \approx
-23.7$ fm), and so a contact interaction---as in the pionless
theory---dominates the $r \rightarrow 0$ behavior of $u_{{\rm {}^1S_0}}(r)$.
Thus in $np \rightarrow d \gamma$ $b=-1$ and $c=-1/4$. The contact
current thus contributes to threshold $np$ capture at $O(Q^{7/4})$
relative to leading.  The enhanced importance of the short-distance
contact current is borne out in explicit calculations of $np
\rightarrow d \gamma$.  It is even more noticeable in calculations of
$nd \rightarrow t \gamma$, and $n {}^3{\rm He} \rightarrow {}^4 {\rm
  He} \gamma$~\cite{Girlanda:2010vm}, where supression of the
one-body piece of the matrix element renders the relative importance of
different two-body currents more transparent.

This enhancement by $1.25$ powers relative to the NDA estimate will
affect any short-distance operator that mediates a ${}^3$S$_1
\leftrightarrow {}^1$S$_0$ transition. In particular, it also occurs
for the solar-fusion process $pp \rightarrow d e^+ \nu_e$, and for
related processes (e.g. muon capture) that proceed via the
Gamow-Teller operator in the NN system. As in the case of threshold
capture, this enhancement also affects the relative importance of
short-distance pieces of two-nucleon operators when the NN system is
embedded in a three- or four-nucleon system that undergoes a weak
transition (cf. Refs.~\cite{Marcucci:2011jm,Marcucci:2013tda,Gazit:2008vm}). We review our results for
NN system processes in Table~\ref{tab:counting}.

This enhancement would be even more dramatic for reactions that
involved the transition ${}^1$S$_0 \rightarrow {}^1$S$_0$, since there
the effect of the short-distance operator increases by two full
orders. (The $\chi$EFT power counting for short-distance operators
becomes the pionless EFT counting in this channel.)
Such transitions occur inside, e.g. ${}^3$He, when electrons scatter from
that nucleus. The analysis of Ref.~\cite{Piarulli:2012bn} could be revisited
in light of this finding, to see if an improved description of the
trinucleon EM form factors results when the anomalous dimensions of
short-distance operators are accounted for.

We stress that RG invariance means that the modifications to the power counting
we discuss here are, to a significant extent, independent of the details of
the cutoff function, or the  numerical value of the cutoff. In particular,
in all contemporary implementations of $\chi$EFT the ${}^1$S$_0$ channel
at LO is dominated by the short-distance potential,
and so the wave function $\psi \sim 1/r$ for distances between
the breakdown scale and the cutoff. 
In the $S=1$ channels the situation is more complicated, since the wave functions
$u$ and $w$ do not seem to behave as $r^{3/4}$ for the distances at which contemporary
$\chi$EFT potentials are regulated~\cite{Ekstrom:2013kea,Epelbaum:2014efa,Epelbaum:2014sza}. 
However, a more careful analysis~\cite{PavonValderrama:2005gu}, shows that
this $r^{3/4}$ behavior is the first term in an expansion for $u$ and $w$ 
which converges up to at least 2 fm, well above the regulator
range used in these potentials. Because of this the scaling derived here
should be relevant for contemporary calculations.  The extent to which the corrections
to $u$ and $w$
computed in Ref.~\cite{PavonValderrama:2005gu} modify the details 
of the RG flow of
short-distance operators is a subject for future work. 

Another avenue for future work is to use 
the numerical behavior of three- (or higher-) body wave functions
at short distance to extend the arguments presented here so as to 
derive the modifications to the NDA power counting for short-range 3N,
4N operators. This exemplifies the power of RG
arguments in nuclear EFT~\cite{Bogner:2003wn,Polonyi:2012ff,Furnstahl:2013oba}. In this work the
principle of RG invariance, applied to NN matrix elements of electroweak currents,
showed the necessity of modifying the counting of short-range
operators in nuclear EFT to account for anomalous dimensions.
The singular nature of the potentials that bind nuclei in both $\chi$EFT
and pionless EFT makes those anomalous dimensions negative,
with the result that NDA underestimates how important such operators are.
This implies a pressing need for revised $\chi$EFT calculations of
electron-deuteron scattering, tri-nucleon form factors, threshold radiative
capture, and weak reactions in few-nucleon systems, for such
enhancements could have important consequences in 
observables.

\begin{acknowledgments}

  D.R.P.  thanks the Theory Group at the University of Manchester
  for hospitality during the inception of this work. We are grateful to Mike Birse and Bingwei Long
  for many discussions on this subject. We also thank
  Harald Grie\ss hammer, Hans-Werner Hammer and Bira van Kolck for
  their input. This
  work was supported by the US Department of Energy under grant
  DE-FG02-93ER-40756.

\end{acknowledgments}

\appendix

%\bibliography{EFT-currents}

\end{document}